\documentclass[12pt]{article}
\usepackage{amsfonts}
\usepackage{url}
\usepackage{epsfig}

\advance\textheight2.8cm        
\advance\topmargin-1.4cm
\advance\textwidth2.6cm
\advance\evensidemargin-2.0cm
\advance\oddsidemargin-2.0cm
\def\gzit#1{{\rm (\ref{#1})}} 			
\def\fct#1{\mathop{\rm #1}}	
\def\re{\fct{Re}}
\def\tr{\fct{tr}}
\def\Cz{\mathbb{C}}
\def\lbeq#1{\begin{equation} \label{#1}} 
\def\eeq{\end{equation}} 
\def\bary{\begin{array}}
\def\eary{\end{array}}

\begin{document}

\begin{center}

{\LARGE \bf A simple hidden variable experiment}

\vfill

\centerline{\sl {\large \bf Arnold Neumaier}}

\vspace{0.5cm}

\centerline{\sl Fakult\"at f\"ur Mathematik, Universit\"at Wien}
\centerline{\sl Nordbergstr. 15, A-1090 Wien, Austria}
\centerline{\sl email: Arnold.Neumaier@univie.ac.at}
\centerline{\sl WWW: \url{http://www.mat.univie.ac.at/~neum}}

\end{center}

\vfill
\hfill June 19, 2007

\vfill
{\bf Abstract.} 
An experiment is described which proves, using single photons only,
that the standard hidden variables assumptions (commonly used to derive 
Bell inequalities) are inconsistent with quantum mechanics.
The analysis is very simple and transparent.
In particular, it demonstrates that a classical wave model
for quantum mechanics is not ruled out by experiments demonstrating the 
violation of the traditional hidden variable assumptions. 

\vfill 

\begin{flushleft} 
\\
\end{flushleft}

Arxiv:0706.0155\\
{\bf 2006\hspace{.4em} PACS Classification}: 42.50.Xa, 
secondary 03.65.Ud

\section{Introduction} \label{s.intro}

It is well-known (see, e.g., {\sc Aspect} \cite{Asp}, 
{\sc Clauser \& Shimony} \cite{ClaS}, {\sc Tittel} et al. \cite{TitBG})
that the foundations of quantum theory can be tested by means of 
optical experiments. A natural question is whether there are simple
tests which can be performed in the classroom. 

Recently, {\sc Hillmer \& Kwiat} \cite{HilK} described a quantum eraser 
experiment which can be performed using very simple equipment. 
In a similar spirit spirit, we describe here and analyze an experimental
setting which allows one to demonstrate with ordinary polarized light
that natural hidden variables assumptions (the same used to derive 
Bell inequalities, cf. {\sc Bell} \cite{Bel}, 
{\sc Pitowsky} \cite{Pit}) are inconsistent with quantum mechanics.
For a recent review of hidden variable theories, see
{\sc Genovese} \cite{Gen}.

In contrast to Bell inequalities which need strongly entangled 
two-photon states to give a contradicition with the quantum predictions,
the experiment suggested here works with arbitrary single-photon states 
and only simple optical equipment. Of course, there is also 
entanglement involved -- not between two photons but between the 
polarization and the spatial degrees of freedom of a single photon.

Moreover, the new experiment provides much sharper predictions than 
traditional Bell inequalities, and its very simple analysis gives new 
insights into the reason for the failure of hidden variable assumptions.

Other papers discussing experiments involving the entanglement of 
single photons include 
{\sc Babichev} et al. \cite{BabAL}, {\sc Bartlett} et al. \cite{BarDSW},
{\sc Beige} et al. \cite{BeiEKW}, {\sc Can} et al. \cite{CanKS},
{\sc van Enk} \cite{vEnk}, {\sc Gerry} \cite{Ger}, 
{\sc Hardy} \cite{Har}, {\sc Ikram \& Saif} \cite{IkrS},
{\sc Hessmo} et al. \cite{HesUHB}, {\sc Kim} \cite{Kim},
{\sc Lee \& Kim} \cite{LeeK}, {\sc Peres} \cite{Per},
{\sc Spreeuw} \cite{Spr}, {\sc Tan} et al. \cite{TanWC},
{\sc Wildfeuer} et al. \cite{WiLD}.
In particular, {\sc Hessmo} et al. \cite{HesUHB} and {\sc Wildfeuer} 
et al. \cite{WiLD} verify experimentally
the prediction of single-particle nonlocality by {\sc Tan} et al. 
\cite{TanWC}, and {\sc Babichev} et al. \cite{BabAL} discusses the
detection loophole for single-particle Bell inequality violation.

\bigskip
{\bf Acknowledgments.} 
Thanks to Erich Dolejsi for creating the figures, to Stefan Ram for 
pointing me to the reference \cite{HilK}, and to Norbert Dragon 
for playing the advocatus diaboli in the discussion on the
newsgroup de.sci.physics in Spring 2007, which motivated me to write 
this paper.

\section{The experiment} \label{s.experiment}

\begin{figure}[htb]
\caption{The hidden variable experiment}
\label{f.hidden}~
\begin{center}
\epsfig{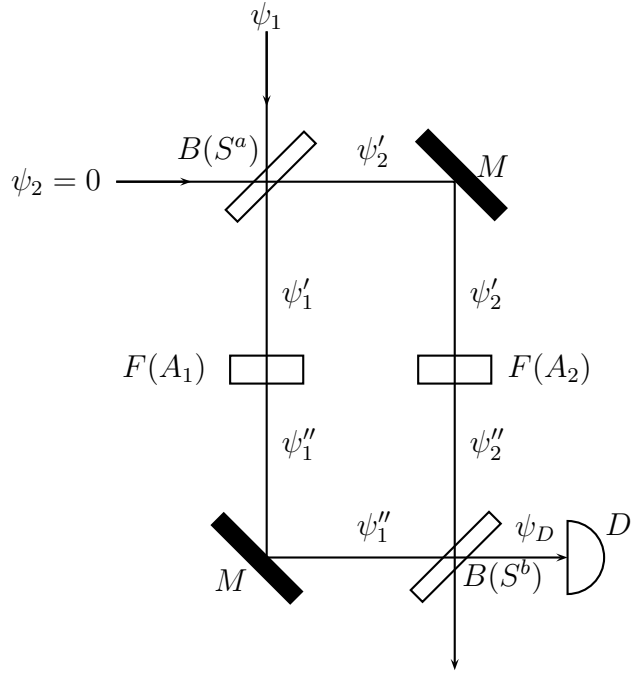}
\end{center}
\end{figure}  

We consider a fixed, symmetric arrangement of optical instruments in 
the form of Figure \ref{f.hidden}. 
For describing the experiment in quantum mechanical terms,
we use unnormalized wave functions $\psi\in\Cz^2$ to denote the state 
of an ensemble of photons in a quasi-monochromatic beam of completely 
polarized light.
$\psi$ is scaled such that $|\psi|^2=\psi^*\psi$ equals the
probability of presence of a photon in the beam in a fixed time 
interval $[t_0,t_1]$; in particular,
dark beams without any photons are described by $\psi=0$.
It is easily checked that optical filters consisting of a combination 
of polarizers are then described by linear transformations of $\psi$
of rank one, and that arbitrary such linear transformations 
$\psi \to A\psi$ ($A=uv^*$) can be realized as long as $|u||v|$ is 
significantly smaller than 1; cf. {\sc Jones} \cite{Jon}.

Each $B(S)$ is a non-polarizing beam splitter with a fixed unitary 
scattering matrix $S\in \Cz^{2\times 2}$,
each $M$ is a mirror, each $F(A)$ is a linear filter transforming the
unnormalized wave function $\psi$ into $A\psi$, with an adjustable 
transformation matrix $A$, and $D$ is a detector registering an 
incident photon with probability $q$. Note that by choosing the 
distances appropriately while keeping the symmetry of the paths, we 
can move the two filters (one under the control of Alice and the other 
under the control of Bob, in the traditional quantum communication 
terminology) as far apart as we like. Thus the experiment can be
given a nonlocal touch, if desired.

The sources of the two beams are not drawn; they are assumed to produce 
completely polarized light described by the unnormalized wave functions 
$\psi_k$ ($k=1,2$). The wave functions are scaled such that initially 
\[
\psi_1^*\psi_1+\psi_2^*\psi_2=1,
\]
corresponding to the presence of just one photon in both beams 
together. The scattering matrices of the two beam splitters are fixed
in the experiment and given by
\[
S^j=\pmatrix{t_1^j & r_2^j \cr r_1^j & t_2^j}~~~(j=a,b),
\]
where $t_k^j$ are the transmission coefficients and $r_k^j$
the reflection coefficients of the two beams; two input beams of the 
beam splitter with wave functions $\psi_1$ and $\psi_2$ are transformed 
into
\[
\pmatrix{\psi_1'\cr \psi_2'} = S^j \pmatrix{\psi_1\cr \psi_2}
= \pmatrix{t_1^j\psi_1+r_2^j \psi_2 \cr r_1^j\psi_1 + t_2^j\psi_2};
\]
cf. {\sc Mandel \& Wolf} \cite[Section 12.12]{ManW}.

We denote by $p(A_1,A_2)$ the probability of detection of a photon 
in the given arrangement, where $A_1$ and $A_2$ are the transformation 
matrices for beam $1$ and beam $2$, respectively, and the second beam
is initially dark.
We analyze the expected dependence of the difference
\lbeq{e.delta}
\Delta(A_1,A_2):=p(A_1,A_2) - p(A_1,0) - p(0,A_2).
\eeq
on the arguments in two ways, first assuming a classical hidden 
variable model, and then assuming quantum mechanics.
By recording enough photons under various settings of $A_1$ and $A_2$,
we can determine $\Delta(A_1,A_2)$, in principle to arbitrary accuracy.
This permits an experimental check on which assumption gives an 
adequate description of the situation.

\bigskip
{\bf Analysis with hidden variables.} 
For the analysis with classical hidden variables, we make the
following assumptions:

(i) The source of beam 1 produces an ensemble of photons which is in 
the classical (but submicroscopic) state $\lambda$ with probability 
density $p(\lambda)$. 

(ii) Whether a photon created at the source in state $\lambda$ reaches 
the detector after passing the $k$th filter depends only on $A_k$ and 
$\lambda$. (This is reasonable since, initially, 
beam 2 is completely dark and hence carries no photons.)

(iii) The conditional probability of detecting a photon which is in 
state $\lambda$ and passes through filter $k$ when $A_k=A$ and 
$A_{3-k}=0$
is $p_k(A,\lambda)$.
 
$p(\lambda)$ and $p_k(A,\lambda)$ are determined by the whole, fixed 
arrangement.
Under these assumptions, the probability of detection of a photon 
when $A_1$ and $A_2$ are arbitrary is
\[
\bary{lll}
p(A_1,A_2) 
&=& \int d\lambda p(\lambda) (p_1(A_1,\lambda)+p_2(A_2,\lambda))\\
&=& \int d\lambda p(\lambda) p_1(A_1,\lambda)+
    \int d\lambda p(\lambda) p_2(A_2,\lambda)\\
&=& p(A_1,0) + p(0,A_2),\\
\eary
\]
hence we get the
\lbeq{e.hidden}
\mbox{\bf hidden variable prediction:~~~~~~}
\Delta(A_1,A_2) =0.
\eeq

\bigskip
{\bf Analysis by quantum mechanics.}
Assuming quantum mechanics, we have, with the notation from the figure,
\[
\psi_1'=t_1^a\psi_1+r_2^a\psi_2,~~~
\psi_2'=r_1^a\psi_1+t_2^a\psi_2,
\]
\[
\psi_1''=A_1\psi_1',~~~
\psi_2''=A_2\psi_2',
\]
\[
\psi_D = t_1^b\psi_1''+r_2^b\psi_2''.
\]
Since the second beam is initially dark, $\psi_2=0$, and we find
\[
\psi_D=t_1^at_1^b A_1\psi_1 + r_1^ar_2^b A_2 \psi_1.
\]
From this, we find
\[
p(A_1,A_2)=q|\psi_D|^2 
=q|t_1^at_1^b A_1\psi_1 + r_1^ar_2^b A_2 \psi_1|^2.
\]
Evaluating \gzit{e.delta} and simplifying, we end up with the
\lbeq{e.quant}
\mbox{\bf quantum prediction:~~~~~~}
\Delta(A_1,A_2)=2q|t_1^at_1^br_1^ar_2^b|^2 \re \psi_1^*A_1^*A_2\psi_1.
\eeq

\section{Discussion} \label{s.disc}

Upon comparing \gzit{e.hidden} and \gzit{e.quant}, we see that
the prediction \gzit{e.quant} of quantum mechanics differ 
significantly from the prediction \gzit{e.hidden} of any hidden 
variable theory satisfying our assumptions. 
The nonlinearity in the squared amplitude formula for 
the probability is responsible for a nontrivial interference term. 
Thus, comparable to destructive interference in two-slit experiments, 
constructive interference is the source for the discrepancy between 
\gzit{e.quant} and \gzit{e.hidden}.
(This is an instance of a more general phenomenon discussed 
by {\sc Malley} \cite{Mal} in a more abstract context, that 
-- under much stronger assumptions --
hidden variables imply the absence of quantum interference terms.)

\bigskip
In terms of the density matrix $\rho=\psi_1\psi_1^*$, the quantum 
prediction can be expressed as
\[
\Delta(A_1,A_2)=2q|t_1^at_1^br_1^ar_2^b|^2 \re \tr(\rho A_1^*A_2).
\]
This relation remains valid if, in place of a pure state $\psi_1$, 
the source produces photons prepared in an arbitrary mixed state $\rho$,
normalized such that $\tr \rho$ equals the mean number of photons in
the fixed time interval $[t_0,t_1]$.

Since the experiment does not involve photon correlation measurements,
the quantum analysis even holds for multiphoton input, provided one 
takes $\rho$ as the effective single-photon density matrix of the
multiphoton state, again normalized such that $\tr \rho$ equals the 
mean number of photons in the fixed time interval $[t_0,t_1]$.

This allows the experiment to be carried out with strong laser light.
In this case, the number of photons is enormous, and the probabilities
turn into essentially deterministic current strengths. Thus performing
the experiment will leave no doubt about the decision for or against
hidden variables, in contrast to the presence of detection loopholes 
in current experiments on local hidden variable theories;
cf., e.g., {\sc Genovese} \cite{Gen} or 
{\sc Babichev} et al. \cite{BabAL}.

The actual performance of the experiment is expected to reproduce the
quantum predictions, thus excluding a theory satisfying our
hidden variable assumptions.

\bigskip
Note, however, that the experiment can be explained by classical 
stochastic Maxwell equations (as discussed in the book by 
{\sc Mandel \& Wolf} \cite{ManW}, upon interpreting the photon 
number detection rate as proportional to the beam intensity.
This is a classical description, not by particles (photons) but by 
waves. 

Indeed, it is well-known (cf. {\sc Weinberg} \cite{Wei}) that Maxwell's 
theory in vacuum can be regarded as the theory of a classical zero mass 
spin 1 photon field, whose quantization (together with that of a 
classical spin 1/2 electron field) gives quantum electrodynamics (QED).

In this light, the present analysis demonstrates that a classical wave 
model for quantum mechanics is not ruled out by experiments 
demonstrating the violation of the traditional hidden variable 
assumptions. 

In particular, this diminishes the role Bell inequality violations
play for investigations the foundations of quantum physics.
From the new perspective gained by the present analysis, the 
traditional hidden variable assumptions therefore only amount to
{\em hidden particle assumptions}, and the experiments demonstrating 
their violation are just another chapter in the old dispute between the 
particle or field nature of light (cf. {\sc Muthukrishnan} et al. 
\cite[p. 20]{MutSZ}), conclusively resolved in favor of 
the field.

\section{General entangled states} \label{s.entangled}

\begin{figure}[htbh]
\caption {State preparation. The first input beam is assumed 
to be dark.}
\label{f.prep}~
\begin{center}
\epsfig{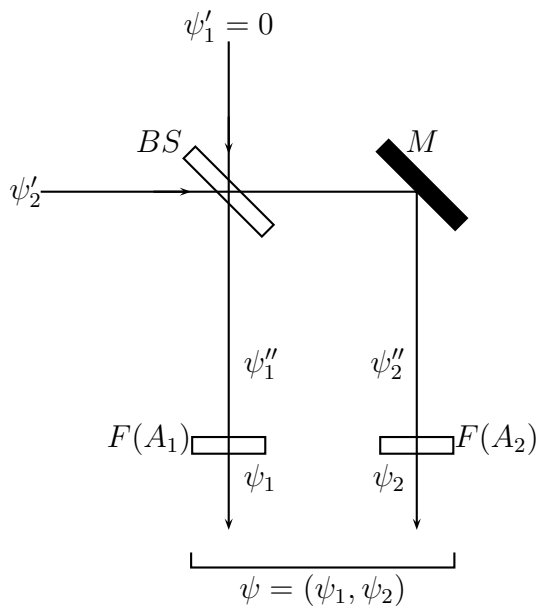}
\end{center}
\end {figure}  

\begin{figure}[htbh]
\caption {State detection. The input state is unknown.
Output entanglement is not measured. Input entanglement can be 
inferred by measuring with different settings of $A_1$ and $A_2$.}
\label{f.meas}~
\begin{center}
\epsfig{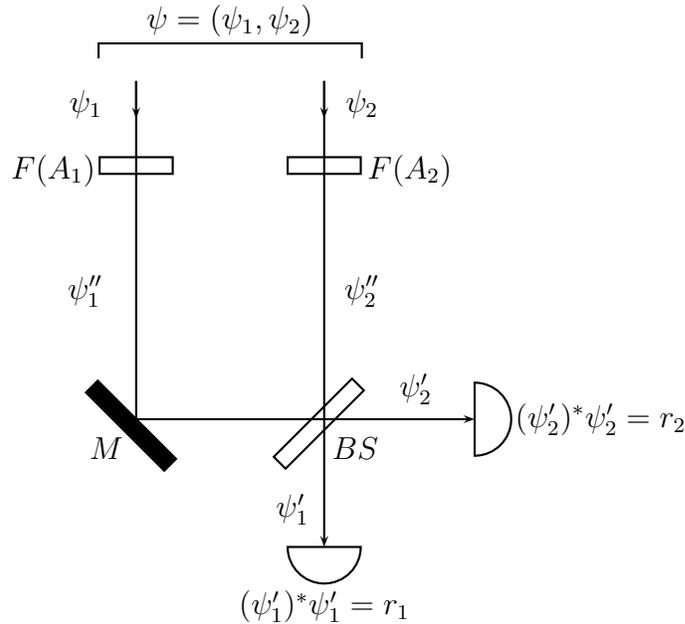}
\end{center}
\end {figure}

It is fairly easy to see that general entangled states can be 
prepared and measured by the arrangements in Figures \ref{f.prep} 
and \ref{f.meas} obtained by splitting our experiment into two halves;
cf. {\sc Kim} \cite{Kim} for alternative preparation and measurement 
settings.

With more beam splitters, through which several narrowly spaced
beams are passed, one can produce a cascade of more complex tensor 
product states. 
Indeed, {\sc Reck} et al. \cite{RecZBB} showed that
(i) any quantum system with only finitely many degrees of freedom can 
be simulated by a collection of spatially entangled beams; 
(ii) in the simulated system, there is for any Hermitian operator $H$ 
an experiment measuring $H$; 
(iii) for every unitary operator $S$, there is an optical arrangement 
in the simulated system realizing this transformation, assuming
lossless beam splitters.

It is not very difficult to show along the lines of \cite{RecZBB} that,
with additional polarizers
and with our convention of scaling state vectors to reflect the 
probability of presence of a photon, one can similarly realize every
subunitary operator $S$, characterized by the condition that all 
eigenvalues of $S^*S$ are bounded by 1.

\bigskip

\end{document}